\documentclass[12pt]{article}
\usepackage{a4wide}
\usepackage{latexsym}
\usepackage{amsmath}
\usepackage{amsfonts}
\usepackage{amscd}
\usepackage{amsthm}
\usepackage{cite}
\usepackage{placeins}

\usepackage{color}

\usepackage{pslatex}
\usepackage{graphicx}
\usepackage[latin1,utf8]{inputenc}
\usepackage[T1]{fontenc}

\allowdisplaybreaks

\newcommand{\bq}{\begin{eqnarray}}
\newcommand{\eq}{\end{eqnarray}}

\theoremstyle{plain}

\begin{document}

\thispagestyle{empty}

\begin{flushright}
  MITP/20-013
\end{flushright}

\vspace{1.5cm}

\begin{center}
  {\Large\bf Correlation functions on the lattice and twisted cocycles \\
  }
  \vspace{1cm}
  {\large Stefan Weinzierl \\
  \vspace{1cm}
      {\small \em PRISMA Cluster of Excellence, Institut f{\"u}r Physik, }\\
      {\small \em Johannes Gutenberg-Universit{\"a}t Mainz,}\\
      {\small \em D - 55099 Mainz, Germany}\\
  } 
\end{center}

\vspace{2cm}

\begin{abstract}\noindent
  {
We study linear relations among correlation functions on a lattice obtained from integration-by-parts identities.
We use the framework of twisted cocycles and determine for a scalar theory a basis of correlation functions,
in which all other correlation functions may be expressed.
   }
\end{abstract}

\vspace*{\fill}

\newpage

\section{Introduction}
\label{sect:intro}

In this article we apply methods and techniques known from perturbative Feynman integral calculations to
non-perturbative lattice correlation functions.

Integration-by-parts identities \cite{Tkachov:1981wb,Chetyrkin:1981qh}
are widely used in perturbative calculations: They relate
Feynman integrals with different powers of the propagators and allow to express 
any Feynman integral from a family of Feynman integrals as a linear combination of basis integrals
(usually called master integrals).
The reduction to master integrals involves (in principle) only linear algebra.
Several publicly available computer programs for Feynman integral reduction
exist \cite{vonManteuffel:2012np,Smirnov:2014hma,Maierhoefer:2017hyi},
which implement the Laporta algorithm \cite{Laporta:2001dd}.
However, for current state-of-the-art calculations the involved systems of linear equations
are rather large and constitute actually a bottleneck.

Quite recently it has been become clear, that Feynman integral reduction can be formulated in the
language of twisted cocycles \cite{Mastrolia:2018uzb,Frellesvig:2019kgj,Frellesvig:2019uqt,Mizera:2019vvs,Mizera:2020wdt}.
This formalism comes with an inner product, given by the intersection number of twisted cocycles.
Thus, if the intersection numbers can be computed efficiently \cite{Weinzierl:2020xyy}
we may bypass the huge systems of linear equations
and compute the reduction to master integrals directly from the inner product.

In this article we would like to point out that the formalism of twisted cocycles has wider
applications and is not restricted to Feynman integral calculations.
The formalism of twisted cocycles has also been applied to perturbative 
scattering amplitudes \cite{Mizera:2017rqa,Mizera:2017cqs,delaCruz:2017zqr,Mizera:2019gea,Mizera:2019blq}
within the Cachazo-He-Yuan formalism \cite{Cachazo:2013gna,Cachazo:2013hca,Cachazo:2013iea}.
Thirdly, we should also mention that concepts closely related to twisted cocycles
in the context of the Batalin-Vilkovisky formalism have been discussed 
in \cite{Schwarz:2008sa,Albert:2008ui,Gwilliam:2012jg,JohnsonFreyd:2012ww}.

In this paper we extend the application of twisted cocycles towards 
correlation functions on a lattice.
At finite coupling these are non-perturbative objects.
On a lattice, the correlation functions are finite-dimensional integrals.
For a scalar theory the dimension of the integrals equals the number of the lattice points.
In this article we focus on a scalar theory.
We determine the dimension and a basis for the twisted cohomology group related to the correlation functions
of a scalar theory on a lattice.
This allows us to find linear relations between different correlation functions at finite coupling on the lattice.
For example, within $\phi^3$-theory we may express a correlation function, where a field occurs at a lattice point $x$ to power two or higher
as a sum of correlation functions, where at each lattice point the field occurs maximally to power one.

In order to avoid misunderstandings let us clearly state that all relations in this article follow from integration-by-parts identities.
We do not consider operator product expansions \cite{Wilson:1972ee}.
The operator product expansion can be used in a continuum space-time 
to express a product of bilocal operators, say at space-time points $x$ and $y$, 
as a linear combination
of regular local operators with coefficients, which contain the short distance singularities.
In this article we always keep the lattice fixed and the lattice spacing $a$ provides an ultraviolet cutoff.

This paper is organised as follows:
In the next section we recall the definition of correlation functions in quantum field theory.
In section~\ref{sect:lattice} we introduce the lattice formulation.
In section~\ref{sect:twisted_cocycles} we define twisted cocycles.
Section~\ref{sect:cohomology} and \ref{sect:reduction} contain the main results of this article:
In section~\ref{sect:cohomology} we determine the twisted cohomology groups for a scalar theory
and in section~\ref{sect:reduction} we present an efficient reduction method for twisted cocycles to
a basis of twisted cocycles.
A few examples are given in section~\ref{sect:examples}.
Finally, section~\ref{sect:conclusions} gives our conclusions.

\section{Correlation functions}
\label{sect:correlation_function}

Fundamental objects in quantum field theory are the $n$-point correlation functions.
They are given in the path-integral formalism by
\bq
 G_n\left(x_1, \dots x_n \right)
 & = &
 \frac{\int {\mathcal D} \phi \; {\phi}(x_1) ... {\phi}(x_n) \; \exp\left(i S\right)}{\int {\mathcal D} \phi \; \exp\left(i S \right)}.
 \nonumber 
\eq
The essential information is given by the path-integral in the numerator, the denominator only provides
the normalisation.
In this article we study the lattice version of integrals of the form
\bq
\label{path_integral}
 \int {\mathcal D} \phi \; \left[{\phi}(x_1)\right]^{\nu_1} ... \left[{\phi}(x_n)\right]^{\nu_n} \; \exp\left(i S\right),
 \;\;\;\;\;\;
 \nu_j \; \in \; {\mathbb N}_0.
\eq
We use lattice regularisation to convert the infinite-dimensional path integral
to a finite-di\-men\-sion\-al integral.
We are interested in determining linear relations among these lattice integrals.
We do this with the help of methods related to twisted cocycles.
As we employ lattice regularisation, all our results are valid for finite coupling
and we never use perturbation theory in this article.

We illustrate the essential points for a scalar theory with Lagrangian
\bq
 {\mathcal L} & = & 
 \frac{1}{2} \left( \partial_\mu \phi \right) \left( \partial^\mu \phi \right) 
 - \sum\limits_{j=2}^{j_{\mathrm{max}}} \frac{\lambda_j}{j!} \phi^j,
\eq
with $\lambda_j \ge 0$ for $2 \le j \le j_{\mathrm{max}}$ and $\lambda_{j_{\mathrm{max}}} \neq 0$.
We call $\lambda_{j_{\mathrm{max}}}$ the leading coupling.
Of particular interest are the cases $j_{\mathrm{max}} \in \{2,3,4\}$.
We call $j_{\mathrm{max}}=4$ a $\phi^4$-theory,
$j_{\mathrm{max}}=3$ a $\phi^3$-theory and
$j_{\mathrm{max}}=2$ a $\phi^2$-theory.
The latter is of course a trivial free theory.

\section{Lattice formulation}
\label{sect:lattice}

As it is standard in lattice formulations we continue to Euclidean time.
We denote by $D \in {\mathbb N}$ the number of space-time dimensions.
A lattice $\Lambda$ with lattice spacing $a$ is specified by a $D$-tuple $(N_0, N_1, \dots, N_{D-1})$, where 
$N_\mu$ gives the number of lattice points in the $\mu$-th direction.
Often it will be convenient to take the same number of points in any direction: $N_0=\dots=N_{D-1}=L$.
We assume periodic boundary conditions.
The lattice has
\bq
 N & = & \prod\limits_{\mu=0}^{D-1} N_\mu
\eq
lattice points.
A lattice point is specified by a $D$-tuple
\bq
 x & = &
 \left( j_0, j_1, \dots, j_{D-1} \right),
 \;\;\;\;\;\;
 0 \; \le \; j_i \; < \; N_i.
\eq
We introduce dimensionless variables
\begin{align}
 \hat{\phi} 
 & = 
 a^{\frac{D-2}{2}} \phi,
 &
 \hat{\lambda}_j
 & = 
 a^{2-\frac{\left(j-2\right)\left(D-2\right)}{2}} \lambda_j.
\end{align}
The field at a lattice point is denoted by
\bq
 \hat{\phi}_x
 & = &
 \hat{\phi}_{j_0, j_1, \dots, j_{D-1}}.
\eq
The Euclidean lattice action $S_E$ is a polynomial in the $N$ variables $\hat{\phi}_x$ and given by
\bq
 S_E
 & = &
 \sum\limits_{x \in \Lambda}
 \left( 
 -  \sum\limits_{\mu=0}^{D-1} \hat{\phi}_x \hat{\phi}_{x+a e_\mu}
 + D \hat{\phi}_x^2
 + \sum\limits_{j=2}^{j_{\mathrm{max}}} \frac{\hat{\lambda}_j}{j!} \hat{\phi}_x^j
 \right).
\eq
$\hat{\phi}_{x+a e_\mu}$ denotes the next lattice point in the (positive) $\mu$-direction modulo $N_\mu$.
On the lattice, the path integral of eq.~(\ref{path_integral}) becomes the $N$-fold integral
\bq
\label{lattice_integral}
 I{\footnotesize \left(\begin{array}{ccc} \nu_1, & \dots, & \nu_n \\ x_1, & \dots, & x_n \end{array} \right)}
 & = &
\int\limits_{{\mathcal C}^N} d^N\hat{\phi} \; \hat{\phi}_{x_1}^{\nu_1} \dots \hat{\phi}_{x_n}^{\nu_n} \; \exp\left(-S_E\right).
\eq
This is a finite-dimensional integral.

\begin{figure}
\begin{center}
\includegraphics[scale=1.0]{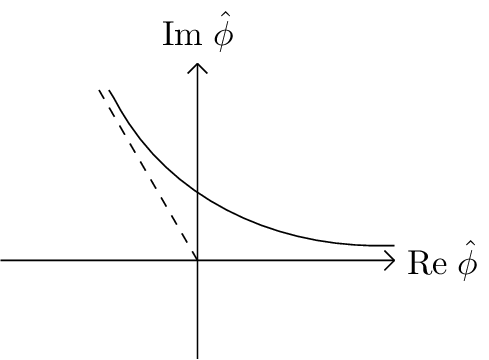}
\end{center}
\caption{
A possible integration contour for $\phi^3$-theory. The asymptotic values are
$\arg \hat{\phi} = 2 \pi /3$ and $\arg \hat{\phi} = 0$.
}
\label{fig1}
\end{figure}

Let us now discuss the integration contour ${\mathcal C}^N$.
We require that each field variable $\hat{\phi}_x$ is integrated along the same curve ${\mathcal C}$ in ${\mathbb C}$ 
and that the integrand vanishes at the boundary.
In the case where $j_{\mathrm{max}}$ is an even number
we may take the standard integration contour 
along the real axis from minus infinity to plus infinity.
However, for $j_{\mathrm{max}}$ odd this is not possible, as the potential is not bounded from below for real field values.
We may get around this by taking an integration contour for a single field to approach for example
$\arg \hat{\phi} = 2 \pi /j_{\mathrm{max}}$ and $\arg \hat{\phi} = 0$ at the boundaries.
For $j_{\mathrm{max}}=3$ this is illustrated in fig.~\ref{fig1}.
In general there are $(j_{\mathrm{max}}-1)$ independent integration contours, specified by the asymptotic values
$\arg \hat{\phi} = 2 \pi j /j_{\mathrm{max}}$ and $\arg \hat{\phi} = 0$  with $1 \le j < j_{\mathrm{max}}$.
Throughout this article we keep the integration contour fixed. 
The precise form of the integration contour does not matter,
the only requirement is that the integrand vanishes on the boundary of the integration contour.
Let us note that the integrand is a holomorphic function of the $N$ variables $\hat{\phi}_x$ on ${\mathbb C}^N$.

\section{Twisted cocycles}
\label{sect:twisted_cocycles}

Let us now reformulate the lattice integrals in the language of twisted cocycles.
We define a function $u$, a one-form $\omega$ and a $N$-form $\Phi$ by
\bq
\label{def_twisted}
 u & = & \exp\left(-S_E\right),
 \nonumber \\
 \omega & = & d \ln u \;= \; -d S_E \; = \; \sum\limits_{x \in \Lambda} \omega_x d\hat{\phi}_x,
 \nonumber \\
 \Phi & = & \hat{\phi}_{x_1}^{\nu_1} \dots \hat{\phi}_{x_n}^{\nu_n} \; d^N\hat{\phi}.
\eq
In terms of these quantities, the integral in eq.~(\ref{lattice_integral}) can be written as
\bq
 I{\footnotesize \left(\begin{array}{ccc} \nu_1, & \dots, & \nu_n \\ x_1, & \dots, & x_n \end{array} \right)}
 & = &
\int\limits_{{\mathcal C}^N} u \; \Phi.
\eq
The one-form $\omega$ defines a covariant derivative $\nabla_\omega=d+\omega$.
By assumption, the integrand vanishes on the boundary of the integration.
We therefore have the integration-by-parts identities
\bq
 0 & = & \int\limits_{{\mathcal C}^N} d^N\hat{\phi} \; \frac{\partial}{\partial \hat{\phi}_x} 
 \left[ \hat{\phi}_{x_1}^{\nu_1} \dots \hat{\phi}_{x_n}^{\nu_n} \; \exp\left(-S_E\right) \right].
\eq
In terms of $\Phi$ this translates  to the statement that the integral is invariant under transformations
\bq
\label{equivalence_relation}
 \Phi' & = & \Phi + \nabla_\omega \Xi,
\eq
for any $(N-1)$-form $\Xi$.
In addition, $\Phi$ is obviously $\nabla_\omega$-closed.
We define the twisted cohomology group $H_\omega^N$ as the equivalence class 
of $\nabla_\omega$-closed $N$-forms
modulo exact ones.
We denote the equivalence classes by $\langle \Phi |$ and refer to these as twisted cocycles.
In a similar way we denote the integration cycle by $| {\mathcal C}^N \rangle$
and refer to it as a twisted cycle.
Our original integral is then written as
\bq 
 \left\langle \Phi \left| {\mathcal C}^N \right. \right\rangle
 & = &
\int\limits_{{\mathcal C}^N} u \; \Phi
 \; = \;
\int\limits_{{\mathcal C}^N} d^N\hat{\phi} \; \hat{\phi}_{x_1}^{\nu_1} \dots \hat{\phi}_{x_n}^{\nu_n} \; \exp\left(-S_E\right).
 \nonumber 
\eq
The essential facts about twisted cohomology groups are \cite{aomoto1975,cho1995,Aomoto:book}:
(i) the twisted cohomology groups $H^N_\omega$ are finite-dimensional and
(ii) there is a a non-degenerate inner product between $H^N_\omega$ and the dual twisted cohomology group
$(H^N_\omega)^\ast=H^N_{-\omega}$ given by the intersection number.

Let $\langle e_1 |, \dots, \langle e_B |$ be a basis of $H^N_\omega$ and let $| d_1 \rangle, \dots, | d_B \rangle$
the dual basis of $(H^N_\omega)^\ast$, satisfying
\bq
 \left\langle e_i | d_j \right\rangle 
 & = & 
 \delta_{i j}.
\eq
We may express $\langle \Phi |$ as a linear combination of the basis $\langle e_i |$:
\bq
\label{reduction_to_basis}
 \left\langle \Phi \right| 
 & = &
 \sum\limits_{i=1}^B c_i \left\langle e_i \right|,
\eq
where the coefficients $c_i$ are given by the intersection numbers
\bq
\label{intersection_numbers}
 c_i & = & \left\langle \Phi \left| d_i \right. \right\rangle.
\eq
Thus we may write our lattice integral as a linear combination of basis integrals:
\bq
 \left\langle \Phi \left| {\mathcal C}^N \right. \right\rangle
 & = &
 \sum\limits_{i=1}^B c_i \left\langle e_i \left| {\mathcal C}^N \right. \right\rangle.
\eq

\section{Dimensions and bases of twisted cohomology groups}
\label{sect:cohomology}

We now determine the dimensions and bases of the twisted cohomology groups.
Let $x_1, \dots, x_N$ label the lattice points. We consider the ideal
\bq
 J & = & \left\langle \omega_{x_1}, \dots, \omega_{x_N} \right\rangle.
\eq
This is a zero-dimensional ideal and the dimension of the vector space
\bq
 {\mathbb C}^N / J
\eq
is finite.
In addition, there is an isomorphism between a basis of ${\mathbb C}^N / J$
and a basis of $H^N_\omega$.
We may construct a monomial basis of ${\mathbb C}^N / J$ as follows:
Let $G=\langle g_1, \dots, g_s \rangle$ be a Gr\"obner basis of $J$.
Then a monomial basis of ${\mathbb C}^N / J$ is given by all monomials not divisible by any $\mathrm{lt}(g_i)$.
Multiplying the monomials by $d^N\hat{\phi}$ gives a basis of $H^N_\omega$.

For the case of $\phi^{j_{\mathrm{max}}}$-theory we find 
\bq
 B & = & \dim H^N_\omega
 \; = \;
 \left(j_{\mathrm{max}}-1\right)^N,
\eq
e.g.
\bq
 \phi^2\mbox{-theory}: & &
 \dim H^N_\omega
 \; = \;
 1,
 \nonumber \\
 \phi^3\mbox{-theory}: & &
 \dim H^N_\omega
 \; = \;
 2^N,
 \nonumber \\
 \phi^4\mbox{-theory}: & &
 \dim H^N_\omega
 \; = \; 
 3^N.
\eq
A basis of $H^N_\omega$ is given by
\bq
\label{def_basis}
 \left\langle e_i \right| \; : \;\;\;
 \left( \prod\limits_{k=1}^N \hat{\phi}_{x_k}^{\nu_k} \right) d^N\hat{\phi},
 & &
 0  \; \le \; \nu_k \; \le \;  j_{\mathrm{max}}-2.
\eq
For $\phi^2$-theory the basis consists of one element
\bq
 \left\langle e_1 \right| 
 & = &
 1 \cdot d^N\hat{\phi}.
\eq
This is not surprising: $\phi^2$-theory is a free theory and all integrals can be reduced
by integration-by-parts identities to a single Gaussian integral.

For $j_{\mathrm{max}} > 2$ the generators $\omega_{x_1}, \dots, \omega_{x_N}$ of the ideal $J$ are
already a Gr\"obner basis with respect to degree lexicographical ordering or 
degree reverse lexicographical ordering.
$\omega_{x_i}$ is given by
\bq
 \omega_{x_i}
 & = &
 - \frac{\hat{\lambda}_{j_{\mathrm{max}}}}{\left(j_{\mathrm{max}}-1\right)!} \hat{\phi}_{x_i}^{j_{\mathrm{max}}-1} + ...,
\eq
where the dots stand for terms of lower total degree.

\section{Reduction to master integrals}
\label{sect:reduction}
 
Given the basis in eq.~(\ref{def_basis})
we would like to express an arbitrary differential $N$-form $\Phi$ of the form as in eq.~(\ref{def_twisted})
as a linear combination 
of the basis as in eq.~(\ref{reduction_to_basis}).
In principle this can be done by computing the intersection numbers in eq.~(\ref{intersection_numbers}).
However, this is impractical. Our main interest is the application where
the number of lattice points $N$ is large and the number of field insertions $n$ in eq.~(\ref{lattice_integral}) is small.
In this case it is more convenient to use eq.~(\ref{equivalence_relation}) repeatedly \cite{Weinzierl:2020xyy}.
For the interacting theories ($j_{\mathrm{max}}>2$)
we may reduce the power $\nu_i \ge (j_{\mathrm{max}}-1)$ of a field at a space-time point $x_i$ as follows:
Let
\bq
 \Phi
 & = &
 \hat{\phi}_{x_i}^{\nu_i} 
 \left( \prod\limits_{\stackrel{k=1}{k \neq i}}^N \hat{\phi}_{x_k}^{\nu_k} \right) d^N\hat{\phi}
\eq
be a representative of $\langle \Phi |$ and consider the $(N-1)$-form
\bq
 \Xi
 & = &
 \frac{\left(j_{\mathrm{max}}-1\right)!}{\hat{\lambda}_{j_{\mathrm{max}}}} \hat{\phi}_{x_i}^{\nu_i-j_{\mathrm{max}}+1} 
 \prod\limits_{\stackrel{k=1}{k \neq i}}^N \hat{\phi}_{x_k}^{\nu_k} d\hat{\phi}_{x_k}.
\eq
Then
\bq
 \Phi' & = & \Phi + \nabla_\omega \Xi
\eq
has at most the power $\nu_i-1$ at the space-time point $x_i$.
The power of the fields at other space-time points may be increased by one.
However, the total degree of all fields is decreased and therefore this algorithm will terminate.

\section{Examples}
\label{sect:examples}

As a simple example let us consider massless $\phi^4$-theory in $D$ space-time dimensions.
We take $\hat{\lambda}_2=\hat{\lambda}_3=0$ and $\hat{\lambda}_4\neq 0$.
Consider
\bq
 I{\footnotesize \left(\begin{array}{c} 4 \\ x \end{array} \right)}
 & = &
 \int\limits_{{\mathcal C}^N} d^N\hat{\phi} \; \hat{\phi}_{x}^{4} \; \exp\left(-S_E\right).
\eq
As integration contour ${\mathcal C}$ we take the real axis.
At the lattice point $x$ the field $\hat{\phi}_{x}$ occurs to power $4$.
In $\phi^4$-theory on the lattice, integration-by-parts identities allow us to express
this correlation function as a linear combination of correlation functions, where at each lattice point
the field occurs maximally to power two. 
Reducing the integral above to master integrals we find
\bq
 I{\footnotesize \left(\begin{array}{c} 4 \\ x \end{array} \right)}
 & = &
 \frac{6}{\hat{\lambda}_4}
 \int\limits_{{\mathcal C}^N} d^N\hat{\phi} \; \left[ 1 + \hat{\phi}_{x} \sum\limits_{\mu=0}^{D-1} \left( \hat{\phi}_{x+a e_\mu} + \hat{\phi}_{x-a e_\mu} - 2 \hat{\phi}_{x} \right) \right] \exp\left(-S_E\right),
\eq
or
\bq
\label{result_example_1}
 I{\footnotesize \left(\begin{array}{c} 4 \\ x \end{array} \right)}
 & = &
 \frac{6}{\hat{\lambda}_4}
 \left\{
   I{\footnotesize \left(\vphantom{\begin{array}{c} 4 \\ x \end{array}} \right)}
   + \sum\limits_{\mu=0}^{D-1} 
     \left[ I{\footnotesize \left(\begin{array}{cc} 1, & 1 \\ x, & x+a e_\mu \end{array} \right)}
   + I{\footnotesize \left(\begin{array}{cc} 1, & 1 \\ x, & x-a e_\mu \end{array} \right)}
   - 2 I{\footnotesize \left(\begin{array}{c} 2 \\ x \end{array} \right)}
 \right] \right\}.
\eq
Here, $I()$ denotes the integral
\bq
 I\left(\right)
 & = &
 \int\limits_{{\mathcal C}^N} d^N\hat{\phi} \; \exp\left(-S_E\right).
\eq
As a second example we consider
\bq
 I{\footnotesize \left(\begin{array}{cc} 1, & 3 \\ x, & y \end{array} \right)} 
 & = &
 \int\limits_{{\mathcal C}^N} d^N\hat{\phi} \; \hat{\phi}_{x} \hat{\phi}_{y}^{3} \; \exp\left(-S_E\right).
\eq
In order to exclude degenerate cases we assume that $x$ and $y$ are separated by at least two lattice units.
The reduction to master integrals yields
\bq
 I{\footnotesize \left(\begin{array}{cc} 1, & 3 \\ x, & y \end{array} \right)} 
 & = &
 \frac{6}{\hat{\lambda}_4}
 \int\limits_{{\mathcal C}^N} d^N\hat{\phi} \; \hat{\phi}_{x} \left[ \sum\limits_{\mu=0}^{D-1} \left( \hat{\phi}_{y+a e_\mu} + \hat{\phi}_{y-a e_\mu} - 2 \hat{\phi}_{y} \right) \right] \exp\left(-S_E\right),
\eq
or
\bq
\label{result_example_2}
 I{\footnotesize \left(\begin{array}{cc} 1, & 3 \\ x, & y \end{array} \right)} 
 & = &
 \frac{6}{\hat{\lambda}_4}
 \sum\limits_{\mu=0}^{D-1} 
 \left[
  I{\footnotesize \left(\begin{array}{cc} 1, & 1 \\ x, & y+a e_\mu \end{array} \right)} 
  + I{\footnotesize \left(\begin{array}{cc} 1, & 1 \\ x, & y-a e_\mu \end{array} \right)} 
  - 2 I{\footnotesize \left(\begin{array}{cc} 1, & 1 \\ x, & y \end{array} \right)} 
 \right].
\eq
Please note that the inverse coupling $\hat{\lambda}_4^{-1}$
appears as a prefactor on the right-hand side of eq.~(\ref{result_example_1}) and eq.~(\ref{result_example_2}).

As a third and more involved example we consider
\bq
 I{\footnotesize \left(\begin{array}{cc} 1, & 9 \\ x, & y \end{array} \right)} 
 & = &
 \int\limits_{{\mathcal C}^N} d^N\hat{\phi} \; \hat{\phi}_{x} \hat{\phi}_{y}^{9} \; \exp\left(-S_E\right).
\eq
In order to exclude degenerate cases we assume that $x$ and $y$ are separated by at least three lattice units.
Furthermore, in order to keep the final expressions to a reasonable size, we present here results for $D=1$ 
space-time dimensions. Results for $D>1$ are easily obtained, but yield longer expressions.
The reduction to master integrals for $D=1$ yields
\bq
\label{result_example_3}
\lefteqn{
 I{\footnotesize \left(\begin{array}{cc} 1, & 9 \\ x, & y \end{array} \right)} 
 = 
 \frac{1296}{\hat{\lambda}_4^4}
 \left[ 
  18 I{\footnotesize \left(\begin{array}{cc} 1, & 1 \\ x, & y \end{array} \right)} 
  - 10 I{\footnotesize \left(\begin{array}{cc} 1, & 1 \\ x, & y+a e_0 \end{array} \right)} 
  - 10 I{\footnotesize \left(\begin{array}{cc} 1, & 1 \\ x, & y-a e_0 \end{array} \right)} 
 \right. 
} & &
 \nonumber \\
 & & \left.
  + I{\footnotesize \left(\begin{array}{cc} 1, & 1 \\ x, & y+2 a e_0 \end{array} \right)} 
  + I{\footnotesize \left(\begin{array}{cc} 1, & 1 \\ x, & y-2 a e_0 \end{array} \right)} 
 \right]
 \nonumber \\
 & & 
 + \frac{648}{\hat{\lambda}_4^3}
 \left[
  16 I{\footnotesize \left(\begin{array}{cc} 1, & 1 \\ x, & y \end{array} \right)} 
  - 8 I{\footnotesize \left(\begin{array}{cc} 1, & 1 \\ x, & y+a e_0 \end{array} \right)} 
  - 8 I{\footnotesize \left(\begin{array}{cc} 1, & 1 \\ x, & y-a e_0 \end{array} \right)} 
  + 4 I{\footnotesize \left(\begin{array}{ccc} 1, & 2, & 1 \\ x, & y & y+a e_0 \end{array} \right)} 
 \right. \nonumber \\
 & & \left.
  + 4 I{\footnotesize \left(\begin{array}{ccc} 1, & 2, & 1 \\ x, & y & y-a e_0 \end{array} \right)} 
  - 2 I{\footnotesize \left(\begin{array}{ccc} 1, & 1, & 2 \\ x, & y & y+a e_0 \end{array} \right)} 
  - 2 I{\footnotesize \left(\begin{array}{ccc} 1, & 1, & 2 \\ x, & y & y-a e_0 \end{array} \right)} 
 \right. \nonumber \\
 & & \left.
  + I{\footnotesize \left(\begin{array}{ccc} 1, & 2, & 1 \\ x, & y-a e_0 & y+a e_0 \end{array} \right)} 
  + I{\footnotesize \left(\begin{array}{ccc} 1, & 1, & 2 \\ x, & y-a e_0 & y+a e_0 \end{array} \right)} 
  - 4 I{\footnotesize \left(\begin{array}{cccc} 1, & 1, & 1, & 1 \\ x, & y-a e_0 & y & y+a e_0 \end{array} \right)} 
 \right]
 \nonumber \\
 & &
 + \frac{108}{\hat{\lambda}_4^2}
 \left[
  4 I{\footnotesize \left(\begin{array}{cc} 1, & 1 \\ x, & y \end{array} \right)} 
  + 3 I{\footnotesize \left(\begin{array}{ccc} 1, & 2, & 1 \\ x, & y & y+a e_0 \end{array} \right)} 
  + 3 I{\footnotesize \left(\begin{array}{ccc} 1, & 2, & 1 \\ x, & y & y-a e_0 \end{array} \right)} 
 \right].
\eq
All relations have been verified numerically by Monte Carlo integrations.

\section{Conclusions}
\label{sect:conclusions}

In this paper we investigated linear relations among correlation functions on a lattice, which have their
origin in integration-by-parts identities.
We formulated the problem in terms of twisted cocycles.
We may think of a twisted cocycle as the integrand of a lattice correlation function without the common
factor $\exp(-S_E)$ and defined up to terms which vanish by integration-by-parts identities.
Mathematically, a twisted cocycle is a twisted cohomology class.
For a scalar theory we determined the dimension and a basis of the twisted 
cohomology group.
In particular we showed that for a $\phi^{j_{\mathrm{max}}}$-theory we may express any lattice correlation function
as a linear combination of lattice correlations functions, where at each lattice point the field $\hat{\phi}$
occurs maximally to the power $(j_{\mathrm{max}}-2)$.

There is no principal obstruction to extend the analysis to Yang-Mills theory on the lattice.
However in practice the determination of the dimension and of a basis for the twisted cohomology groups will
be more challenging.
For a scalar theory we profited from the fact that a Gr\"obner basis for the ideal $J$ is easily found
with respect to degree lexicographical ordering or 
degree reverse lexicographical ordering.
This can be traced back to the leading coupling term in the potential
$\frac{\lambda_{j_{\mathrm{max}}}}{j_{\mathrm{max}}!} \phi^{j_{\mathrm{max}}}$.
For Yang-Mills theory on the lattice, the Euclidean action is given as a sum over plaquettes
and a plaquette expands into a product of fields at neighbouring lattice points, and not at the same lattice point.
This makes the determination of the dimension and of a basis for the twisted cohomology groups more challenging.

\subsection*{Acknowledgements}

I would like to thank the anonymous referee 
for bringing references \cite{Schwarz:2008sa,Albert:2008ui,Gwilliam:2012jg,JohnsonFreyd:2012ww} to my attention,
where ideas of homological perturbation theory in the context
of the Batalin–Vilkovisky formalism are discussed.
As this article provides a bridge between 
Feynman integrals and correlation functions on a lattice,
we may transfer ideas of homological perturbation theory to integration-by-parts reduction of Feynman integrals.
This will be an interesting project for the future.

\bibliography{/home/stefanw/notes/biblio}
\bibliographystyle{/home/stefanw/latex-style/h-physrev5}

\end{document}